\documentclass[twocolumn,
spreprintnumbers,amsmath,amssymb]{revtex4}
\usepackage[dvips]{graphics}
\usepackage{color}

\newcommand{\ee}{\end{eqnarray}}
\newcommand{\be}[1]{\begin{eqnarray} \mbox{$\label{#1}$} }

\newcommand{\mtrx}[2]{\left(\begin{array}{#1} #2 \end{array}\right)}
\newcommand{\eref}[1]{(\ref{#1})}
\newcommand{\der}{\mathrm{d}}

\newcommand{\sch}{Schr\"odinger{} }
\newcommand\ie {{\it i.e. }}
\newcommand\eg {{\it e.g. }}

\newcommand{\ul}\underline

\newcommand{\pr}{^\prime}

\renewcommand{\vec}[1]{\text{\boldmath{$ #1 $}}}

\newcommand{\idm}{\mathbf{1}}

\newcommand{\zc}{{z^*}}

\newcommand{\Hss}{\mathcal{H}}
\newcommand{\Qss}{\mathcal{Q}}

\begin{document}

\title{Pedestrian index theorem \`a la Aharonov-Casher \\ for bulk threshold modes in corrugated multilayer graphene}

\author{J. Kailasvuori}
\email {kailas@physik.fu-berlin.de}

\affiliation{Institut f\"ur theoretische Physik,
Freie Universit\"at Berlin, Arnimallee 14, 14195 Berlin,
Germany\\
Max-Planck-Institut f\"ur Physik komplexer Systeme, N\"othnitzer Str. 38, 01187 Dresden, Germany }

\date{April 24, 2009}

\begin{abstract}
\noindent
Zero-modes, their  topological degeneracy and relation to index theorems have attracted attention in the study of single- and bilayer graphene. For negligible scalar potentials, index theorems can explain why the degeneracy of the zero-energy  Landau level of a Dirac hamiltonian is not lifted by gauge field disorder, for example due to ripples, whereas other Landau levels become broadened by the inhomogenous effective magnetic field. That also the bilayer hamiltonian supports such protected bulk zero-modes was proved formally by Katsnelson and Prokhorova to hold on a compact manifold  by using the Atiyah-Singer index theorem. Here we  complement and generalize this result in a pedestrian way by pointing out that the simple argument  by Aharonov and Casher  for degenerate zero-modes of a Dirac hamiltonian in the infinite plane extends naturally to the multilayer case. The degeneracy remains, though at nonzero energy,  also in the presence of a gap. These threshold modes make the spectrum asymmetric. The rest of the spectrum, however, remains  symmetric even in arbitrary gauge fields, a fact related to supersymmetry. Possible benefits of this connection are discussed. 

\end{abstract}

\maketitle

\section{Introduction}
\noindent
Since the experimental realization of graphene it has become clear that a suspended sheet of graphene is not flat but corrugates into a rippled structure.\cite{Meyer, Scott} In the tight binding-model for graphene these ripples with their intrinsic curvature lead to a local modification of the hopping amplitudes. In the low-energy limit given by a Dirac hamiltonian the ripples enter as an effective disorder potential, of which the vector part can be interpreted as a nonuniform effective magnetic field.\cite{KanMel} The impact of this disorder potential on the spectrum and on transport properties has attracted a lot of interest (see ref.~\onlinecite{Cas}). For example, ref.~\onlinecite{GuiKatVoz} studies numerically the low-energy spectrum in the presence of certain ripple configurations and finds it to be considerably changed when the effective magnetic length is comparable to the ripple size.  Zero-energy Landau-level-like states  can then exist within one ripple and their degeneracy  is no lifted by the inhomogeneity of the effective magnetic field.  This should be observable as a peak at zero energy  in the density of states.\cite{GuiKatVoz, Weh}  In presence of a scalar potential the degeneracy is lifted, which can also be seen in   ref.~\onlinecite{GuiKatVoz}.  

In the quantum Hall problem the small effective magnetic field due to ripples is combined with the strong uniform external magnetic  field into a total nonuniform magnetic  field.  It is observed that the zero-energy Landau level in graphene remains strongly peaked, whereas the other Landau levels are broadened, possibly due to  the inhomogenous field caused by the ripples.\cite{Gie}   

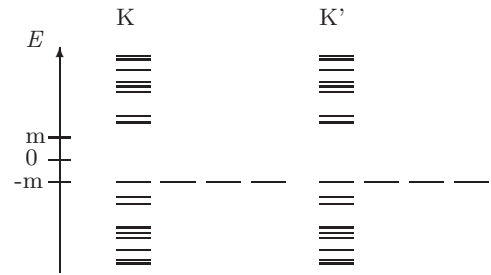
\begin{figure}[ht]
\setlength{\unitlength}{1.5mm}
\begin{picture}(0,23)(12,-10)

\put(-5,-10){\vector(0,1){20}}
\put(-8,10){\textit{E}}
\put(-8,-0.5){0}
\put(-6,0){\line(1,0){2}}

\put(-8,1.5){m}
\put(-6,2){\line(1,0){2}}
\put(-9,-2.5){-m}
\put(-6,-2){\line(1,0){2}}

\put(0,12){K}
\put(18,12){K'}

\multiput(0,0)(18,0){2}{

\multiput(0,-2)(4,0){4}{\line(1,0){3}}

\put(0,3.3){\line(1,0){3}} 
\put(0,3.9){\line(1,0){3}} 
\put(0,6){\line(1,0){3}} 
\put(0,6.5){\line(1,0){3}} 
\put(0,6.9){\line(1,0){3}} 
\put(0,8){\line(1,0){3}} 
\put(0,8.9){\line(1,0){3}} 
\put(0,9.2){\line(1,0){3}} 

\put(0,-3.3){\line(1,0){3}} 
\put(0,-3.9){\line(1,0){3}} 
\put(0,-6){\line(1,0){3}} 
\put(0,-6.5){\line(1,0){3}} 
\put(0,-6.9){\line(1,0){3}} 
\put(0,-8){\line(1,0){3}} 
\put(0,-8.9){\line(1,0){3}} 
\put(0,-9.2){\line(1,0){3}} 
}

\end{picture}\\
\caption{An imagined typical spectrum for  electrons in graphene with a gap and a random magnetic field due to ripples.  (The horizontal position carries no meaning.) The two valleys K and K' are time-reversed copies in absence of a real magnetic field. For  ripples without any spatial symmetries the spectrum of  each valley will in general be non-degenerate and of random spacing, except for the threshold modes at $|E|=m$ . Their degeneracy depends only on the total effective flux $\Phi^\textrm{K}=-\Phi^\textrm{K'}=4.1\phi_0$. The rest of the spectrum is symmetric around $E=0$ due to supersymmetry. }
\label{f:sym1}
\end{figure}

The stability of  zero-energy states---{\it zero-modes}--- of a Dirac electron in a magnetic field of arbitrary shape is  understood as a consequence of index theorems, which relate analytical properties of operators to topological properties of the space and the fields involved.  (See \eg refs.~\onlinecite{Nak, NieSem, Tha}) There are several index theorems, with different range of applicability. Maybe the most famous one is the Atiyah-Singer index theorem\cite{AtiSin}, that applies to elliptic differential operators on a compact manifold of even dimension, \eg the sphere or the torus. It 
 states that for  elliptic differential operators with the Fredholm property (\eg Dirac operators $\Pi_\pm=\Pi_x\pm i \Pi_y$ on a torus, with $\vec{\Pi}=-i\nabla+\vec{A}$), the analytical index   (the number of zero-modes of $\Pi_+$, \ie number of solutions to $\Pi_+ u(\vec{x})=0$, minus the number of zero-modes of $\Pi_-$) equals the topological index (\ie the total magnetic flux of the gauge field, a topological integer according to the Dirac monopole quantization condition on compact manifolds). 
 In some cases it has been possible to find index theorems on non-compact manifolds or for odd space dimensions.  One of many reason for  generalizations to be interesting is that  many Dirac operators occurring in quantum mechanics (in particular Dirac operators  on the infinite plane to be studied here) usually do not have the Fredholm property, a prerequisite for the Atiyah-Singer theorem.  See refs.~\onlinecite{ Nak, NieSem, Tha} for a review and references  on these generalizations. 
 
Fortunately for physicists the essence of  this beautiful and powerful but rather advanced mathematics is manifested in some simple examples requiring only elementary  quantum mechanics, many of them living on infinite manifolds.  A nice feature of these examples is that the wave functions of the zero-modes are obtained explicitly.  
Jackiw and Rebbi\cite{JacReb} found that in a 1d Dirac theory there could be a topologically protected zero-mode localized at a mass soliton.  
 Aharonov and Casher\cite{AhaCas} found a very short and simple argument for zero-modes of massless Dirac electrons in a 2d plane in a nonsingular perpendicular magnetic field of arbitrary shape but finite range. The degeneracy of the zero-modes is only determined by the total flux, just like in the Atiyah-Singer theorem. In the presence of a mass  $m$ these zero-modes turn into degenerate {\it threshold modes}. Depending on the sign of the total flux, they  sit either at the $E=+m$ or at the $E=-m$ threshold of the gapped spectrum.  
 
In this paper we focus on Aharonov's and Casher's pedestrian argument.  In section~\ref{s:multi} we point out how simply it also extends to some of the hamiltonians that have been considered for multilayer graphene.  Stability of zero-modes in  rippled bilayer graphene was already considered by Katsnelson and Prokhorova\cite{KatPro}, there in the general but abstract language of the Atiyah-Singer theorem, thus applying to a compact manifold, in this case the torus resulting from periodic boundary conditions.  Our result is complementary by applying to the infinite plane. It also offers a simple generalization to multilayers. The pedestrian argument goes beyond the Atiyah-Singer theorem by showing that the zero-modes are present also  when the effective flux though the graphene sheet is not an integer. Finally, our argument is of pedagogic value as it does not require any higher mathematics and also gives concrete wave-functions. 
 
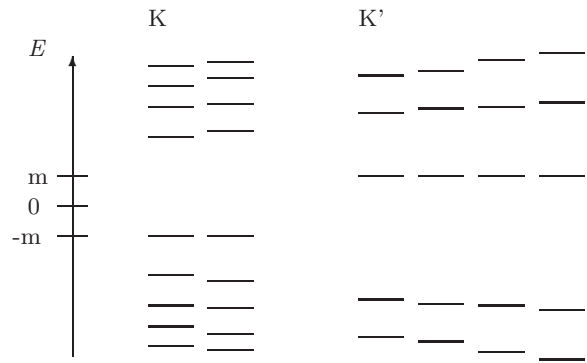
\begin{figure}[ht]
\setlength{\unitlength}{2mm}
\begin{picture}(0,22.5)(10,-10)

\put(-5,-10){\vector(0,1){20}}
\put(-8,10){\textit{E}}
\put(-8,-0.5){0}
\put(-6,0){\line(1,0){2}}

\put(-8,1.5){m}
\put(-6,2){\line(1,0){2}}
\put(-9,-2.5){-m}
\put(-6,-2){\line(1,0){2}}

\put(0,12){K}
\put(14,12){K'}

\put(0,0){
\multiput(0,-2)(4,0){2}{\line(1,0){3}}

\put(0,4.6){\line(1,0){3}} 
\put(4,5){\line(1,0){3}} 
\put(0,6.6){\line(1,0){3}} 
\put(4,6.8){\line(1,0){3}} 
\put(0,8){\line(1,0){3}} 
\put(4,8.5){\line(1,0){3}} 
\put(0,9.3 ){\line(1,0){3}} 
\put(4,9.6){\line(1,0){3}}

\put(0,-4.6){\line(1,0){3}} 
\put(4,-5){\line(1,0){3}} 
\put(0,-6.6){\line(1,0){3}} 
\put(4,-6.8){\line(1,0){3}} 
\put(0,-8){\line(1,0){3}} 
\put(4,-8.5){\line(1,0){3}} 
\put(0,-9.3 ){\line(1,0){3}} 
\put(4,-9.6){\line(1,0){3}}

}

\put(14,0){
\multiput(0,2)(4,0){4}{\line(1,0){3}}

\put(0,6.2){\line(1,0){3}} 
\put(4,6.5){\line(1,0){3}} 
\put(8,6.6){\line(1,0){3}} 
\put(12,6.9){\line(1,0){3}} 
\put(0,8.7){\line(1,0){3}} 
\put(4,9 ){\line(1,0){3}} 
\put(8,9.7){\line(1,0){3}} 
\put(12,10.2){\line(1,0){3}} 

\put(0,-6.2){\line(1,0){3}} 
\put(4,-6.5){\line(1,0){3}} 
\put(8,-6.6){\line(1,0){3}} 
\put(12,-6.9){\line(1,0){3}} 
\put(0,-8.7){\line(1,0){3}} 
\put(4,-9 ){\line(1,0){3}} 
\put(8,-9.7){\line(1,0){3}} 
\put(12,-10.2){\line(1,0){3}} 

}

\end{picture}\\
\caption{Same sketch as in Fig.~\ref{f:sym1} but now including a real magnetic field with $\Phi_\textrm{EM}=-3.2\phi_0$. (The horizontal position only used to guide the eye.)
The quasi Landau levels  are broadened by the ripples, except the Landau level of the threshold modes.  However, with $\Phi_\textrm{rip}=1.1\phi_0$ and therefore  $\Phi^{\textrm{K}}=\Phi_\textrm{EM}+\Phi_\textrm{rip}=-2.1 \phi_0 $ and $\Phi^{\textrm{K}\pr}=\Phi_\textrm{EM}-\Phi_\textrm{rip}= -4.3 \phi_0$ the threefold degeneracy of the threshold modes at $E=-m$ ($E=+m$) of valley K (K')  becomes twofold (fourfold).  The rest of the spectrum is in each valley symmetric about $E=0$ due to supersymmetry.    
  }
\label{f:sym2}
\end{figure}

In section~\ref{s:threshold} we also extend the discussion by  including a mass term. In a single layer this corresponds to breaking the sublattice symmetry of the bipartite honeycomb lattice, like in experiments\cite{Eli} on hydrogenated single-layers. In a multilayer a gap can also been introduced by a transverse potential, like in experiments\cite{Tai, Oos} on gated bilayers.   
A mass term turns the zero-modes into degenerate threshold modes sitting at the gap energy (Figs.~\ref{f:sym1} and \ref{f:sym2}).  In the quantum Hall problem such a term would split the sharp zero-energy Landau-level peak into  two sharp peaks,  symmetrically shifted around zero and related to the two valleys (Fig.~\ref{f:sym2}). Because of the valley degeneracy breaking combination of ripples and an external magnetic field the two peaks would be of different sizes. Such a splitting has actually been observed experimentally\cite{Gie, Gie2009}, but has been attributed to other mechanisms.  
 
A known but often not mentioned point is that all the mentioned topological arguments for a sharp peak at zero energy or at a threshold energy seem to fail if scalar potentials, for example induced by ripples or by impurities, are not negligible. This important caveat is brought up in section~\ref{s:threshold}, but  will remain an open question both for single-layer and multilayer graphene.   

  The third part of the paper makes a note on the symmetry of the spectrum, as illustrated in Figs.~\ref{f:sym1} and \ref{f:sym2}. Apart from the threshold modes, the spectrum of each valley remains symmetric around the zero of energy also in presence of both a nonuniform vector potential and a mass term, provided the mass is constant and that the scalar potential is zero. This symmetry in not due to the chiral symmetry $\sigma_z H_{m,A_0,\vec{A}}\sigma_z =- H_{ -m,-A_0,\vec{A}}$ within each valley, which is only enough to explain this symmetry in the massless case for example studied in ref.~\onlinecite{GuiKatVoz}.  The more general reason of this symmetry we relate instead to the supersymmetric quantum mechanics formulated by Witten\cite{Wit}. Further benefits of this remark could come from connecting to the rich literature on analytic results based on supersymmetry, for example on scattering of Dirac electrons in slowly decreasing magnetic fields in which asymptotic states are difficult to define. Since the multilayer hamiltonians also have the supersymmetric structure, we expect that many such analytic results  should have analogs in the multilayer case.

 \section{Threshold modes in rippled graphene}\label{s:threshold}
\noindent
Assume a general magnetic field 
 $B(x,y)=\partial_x A_y-\partial_y A_x$   of compact support, \ie there is a finite disk outside which  $B(\vec{x})=0$. Define the Dirac operators $\vec{\Pi}=-i\nabla+\vec{A}$ and $\Pi_\pm=\Pi_x\pm i \Pi_y$.  We set $\hbar=e=1$. The argument by Aharonov and Casher, that we review in the appendix,   shows that a massless 2d Dirac hamiltonian
\be{1layer}
H=
v\mtrx{cc}
{0 & \Pi_- \\ \Pi_+ & 0}\, ,
\ee
  has $n$ zero-modes, where $n\geq 0$ is only determined by the total flux $\Phi=\int \der^2 x\, B= \pm \phi_0 (n+\epsilon)$  (with $\phi_0= h/e$ and $0<\epsilon \leq 1$) and hence independent of the shape of the magnetic field. In the presence of a mass term $\delta H = m\sigma_z$ these $n$ zero-modes turn into $n$  threshold modes, either with energy $E=+m$ or with $E=-m$, depending of the sign of the flux. 
These solutions are of the form 
\be{tmodes}
\begin{array}{rcl}
\psi_{+m} &=& \mtrx{c}{u \\ 0 }\\
u(\vec{x}) &=& f(z)e^{W(\vec{x})}  \\
\Pi_+ u & = & 0 
\end{array}
\hspace{0.5cm}
 \textrm{or}
 \hspace{0.5cm}
 \begin{array}{rcl}
\psi_{-m} &=& \mtrx{c}{0 \\ v }\\
v(\vec{x}) &=& f(z^*)e^{-W(\vec{x})}  \\
\Pi_- v & = & 0 
\end{array}
\ee
($z= (x_1+ix_2)/2$). $f$ has to be a polynomial for single-valuedness and regularity. The magnetic field enters into $W(\vec{x})=\frac{1}{\phi_0}\int \der^2 x^\prime \, 
B(\vec{x^\prime}) \ln |\vec{x}-\vec{x^\prime}|$, which far away from the region with the magnetic field behaves asymptotically as $e^{\pm W(\vec{x}) }\sim |\vec{x}|^{\pm \Phi/\phi_0}$.   Normalizabilty requires the exponent to be negative and requires $f$ to be maximally of degree $n-1$.  This gives $n$ linearly independent polynomials and hence $n$ threshold modes, which are zero-modes in the massless limit.

Note that in the case of $|\Phi | / \phi_0=n+1=\tilde{n}$ integer flux quanta,  
 the  $\tilde{n}^{\mathrm{th}}$ solution  $u\propto z^{\tilde{n}-1}|\vec{x}|^{-\tilde{n}}$ is not strictly square integrable in the plane  since the integral diverges logarithmically. On a compact manifold this "marginal" mode becomes normalizable, therefore the $\tilde{n}=\Phi/\phi_0$-fold degeneracy of the lowest Landau level on a torus.

 The valley K of graphene is described by the hamiltonian \eref{1layer} with $v \approx 10^6\, \mathrm{m/s}$, complemented with the scalar potential $\delta H=-\idm A_0$ and here possibly with the addition of a 
 sublattice symmetry breaking term $\delta H = m\sigma_z$. The potential $m(\vec{x})$ acts with opposite signs on the two inequivalent orbitals A and B of the bipartite 
  honeycomb lattice. The components of the  spinor
   $(u, \,  v)^\mathrm{T} = (\psi_\mathrm{A}^{\mathrm{K}},  \, \psi_\mathrm{B}^{\mathrm{K}})^\mathrm{T}$ 
    refer to these 
 orbitals. The gauge field $A_\mu$ ($\mu=0,1,2$) is here a sum of the  electromagnetic part $A_{\mu,\textrm{EM}}$ and an effective field $A_{\mu,\textrm{rip}}$ due to ripples.\footnote{One could also include a {\it  inter}-valley disorder potentials, but this is here  neglected assuming that the  characteristic wave-number of the disorder is much smaller than the wave-number of the Dirac points. } 
 The corresponding hamiltonian  for valley K'  is $H^{\mathrm{K}\pr}=-v\vec{\sigma}\cdot(-i\nabla+\vec{A}_\textrm{EM}-\vec{A}_\textrm{rip})-\idm A_0-m\sigma_z$ acting on   $(\psi_\mathrm{B}^{\mathrm{K\pr}},  \, \psi_\mathrm{A}^{\mathrm{K}\pr})^\mathrm{T}$.
Thus, the effective magnetic field due to ripples changes sign, but not  the external magnetic field. In absence of an external magnetic field the two valleys are mapped onto each other under time-reversal $\sigma_x( H^{\mathrm{K}})^*\sigma_x=H^{\mathrm{K}\pr}$, thus preserving the time-reversal symmetry of the total graphene hamiltonian also in presence of ripples.  With both $\vec{A}_\textrm{EM}$ and $ \vec{A}_\textrm{rip} $ nonzero the degeneracy between the two valleys is lifted. 

In this context we note that in the absence of a scalar potential but with a nonzero gap term  $m\sigma_z$ the threshold modes of the two valleys sit at the same energy  if $|\Phi_\textrm{rip}|>|\Phi_\textrm{EM}|$. For example, with $\Phi_\textrm{rip}>0$ the threshold modes  of both valleys  sit at $E=-m$.   In the time-reversal invariant case $\vec{A}_\textrm{EM}=0 $ (Fig.~\ref{f:sym1}) both valleys count the same number of threshold modes  and they are related by time-reversal $\psi_{-m}^{\textrm{K}\pr}=\sigma_x(\psi_{-m}^{\textrm{K}})^* $. In the opposite case  $|\Phi_\textrm{rip}|<|\Phi_\textrm{EM}|$, including the quantum Hall scenario, the threshold modes of the two valleys sit at opposite energies (Fig.~\ref{f:sym2}). Only in the special case $\Phi_\textrm{rip}=0$ are they equally many. In general, in a quantum Hall problem with a sublattice symmetry breaking gap $m\sigma_z$, the zero energy Landau-level peak in the density of states is split into two symmetrically shifted peaks centered at $E=+m$ and $E=-m$, respectively, but of different sizes due to the different fluxes $\Phi^\textrm{K}=\Phi_\textrm{EM}+\Phi_\textrm{rip}$ and $\Phi^{\textrm{K}\pr}=\Phi_\textrm{EM}-\Phi_\textrm{rip}$. Experiments\cite{Gie, Gie2009} find at low temperatures the zero-energy Landau level peak to be split into two peaks, resulting in a plateau in the magnetoresistance at zero doping.  We find it suggestive that the two peaks appear to be of different sizes.  However, the splitting was attributed to counterpropagating edge channels dominating the longitudinal resistivity\cite{Aba} or  to Zeeman splitting in the real spin\cite{Zha}.

The effective vector potential derived from the rippling is given by $\vec{A}_\textrm{rip}=g_2 (u_{xx}-u_{yy},2u_{xy}) $ \cite{SuzAnd} in terms of the strain tensor $u_{ij}\equiv \frac{1}{2}(\partial_iu_j+\partial_ju_i+(\partial_ih)(\partial_j h))$ with $\vec{u}(\vec{x})$ and $h(\vec{x})$ being the in-plane and out-of-plane distortions, respectively. $g$ is a coupling parameter that depends on the properties of the bonds.   Observe that $\vec{A}$ does not change under $h\rightarrow -h$. A buckled-up ripple gives therefore  a flux of the same sign as the equivalent buckled-down ripple. Thus, the fluxes of the ripples always add up. The flux of one ripple of length $l$ and height $h$ can be estimated to be  $|\Phi|/\phi_0 \sim h^2/al $, with $a$ being the bond length of the honeycomb lattice.\cite{And} Index theorems, or the  explicit solution found by Aharonov and Casher, explain why ripples can lead to degenerate zero-modes or threshold modes. 

The accompanying effective scalar potential $\delta H = \idm g_1( u_{xx}+u_{yy})$ destroys this exact degeneracy (as it is nonuniform).  An approximate degeneracy could still hold if $g_1$ is small enough. However, it is far from obvious that this is the case for monolayer graphene, considering estimates\cite{SuzAnd} for carbon nanotubes with $g_1\approx 30\, \textrm{eV}$ and $g_2\approx 1.5\, \textrm{eV}$. On the other hand, it is not impossible that the bare value of the scalar potential could be substantially reduced for example through screening. As far as we know this has not been clarified. It   remains therefore unclear  to us if the observed sharpness of the zero energy Landau level is really due to index theorems, with some mechanism suppressing the scalar potential, or if it must be attributed to some other reason.  As for multilayers, to which we now turn to, we do not know of any such estimates. With the above caveat mentioned, we now go on to analyze threshold modes  in multilayers for the case that the scalar potentials would turn out to be negligible there.

\section{Threshold modes in multilayer graphene}\label{s:multi}
\noindent     
The intense study of monolayer graphene has been followed by that of  tight-binding models for bilayer and multilayer graphene, with low-energy continuum models similar to the monolayer one.\cite{MccFal, GuiCasPer, KosAnd, MinMac} These multilayer sheets have exotic properties of their own, but offer also an interesting interpolation back to the source material  graphite. 
In the simplest tight-binding model, the electrons at the  valley K  in a bilayer are in a low-energy range described by the effective hamiltonian 
\be{2layer}
H_\textrm{eff} \propto 
\mtrx{cc}
{0 & \Pi_-^J \\ \Pi_+^J & 0}\, .
\ee
with $J=2$.\cite{MccFal} This hamiltonian acts on the spinor $(\psi_\mathrm{A},\psi_\mathrm{\tilde{B}})$ in a bilayer arranged so that the B-orbitals of the AB-layer are placed on top of the \~A-orbitals in the \~A\~B-layer, giving the leading inter-layer tunneling channel. The superposed orbitals \~A and B dimerize in the low-energy limit, leaving effectively the A and \~B orbitals.     

For a three layers and beyond \cite{GuiCasPer, KosAnd, MinMac} there are several ways of stacking the layers, for example the Bernal stacking (ababa...)---the most common form in natural graphite---or the rhombohedral stacking (abcabc...), with a, b, and c denoting the three different placements of honeycomb sheets that can  occur in a stacking. With the simplest approximation for inter-layer tunneling one obtains for a rhombohedrally stacked $N$-layer the effective hamiltonian \eref{2layer} with $J=N$. For a  Bernal stacked $N$-layer one finds instead 
\be{Nlayer}
H_\textrm{eff}\sim \bigotimes_{J_i} H_{J_i}\, , \hspace{1cm} \sum_i J_i=N
\ee
with  $J_i=2$, except $J_1=1$ if $N$ is odd---thus a tensor product of bilayer and monolayer hamiltonians. Other stackings give a structure that is intermediate between the rhombohedral and the Bernal one.\cite{MinMac}

Also multilayer graphene sheets form ripples\cite{MeyGei}, and  the existence and stability of zero-modes is interesting to address. 
Particularly, Katsnelson and Prokhorova\cite{KatPro} gave a formal proof, based on the Atiyah-Singer index theorem complemented with a result\cite{Pal} on indices of composed elliptical operators. They showed that the mentioned bilayer hamiltonian  on a compact 2d manifold (in particular the torus obtained from assuming periodic boundary conditions) has zero-modes, precisely twice as many as the monolayer hamiltonian \eref{1layer}. The contribution of the present paper is to note in a pedestrian way (although now for an infinite plane)  that this fact and the generalization to arbitrary $J $  follow from the argument of Aharonov and Casher with the following straightforward extension. 
Assume $\Phi$ to be negative. There are the  $n$ zero-modes $(u, \, v)^\textrm{T}=(  f(z)e^{W(\vec{x})}, 0)^\textrm{T}$ that satisfy $\Pi_+ fe^W=0$. Since $\Pi_+=-i\partial_{z^*}+A_+$ it follows that
\be{}
\Pi_+  (\zc )^j fe^W  =  -ij  (\zc )^{j-1} fe^W\, .
\ee
As a consequence, all $j=0,\ldots, J-1$ give wave functions  $u=(\zc )^j f(z)e^{W(\vec{x})}$ that are zero-modes of $\Pi_+^J  u=0$.  The hamiltonian \eref{2layer} has therefore $nJ$ zero-modes and the total hamiltonian \eref{Nlayer} has $nN$ zero-modes, independently of the shape of the gauge field and independent of the stacking configuration.  In the presence of a constant mass term  $m\sigma_z$ in the hamiltonian, these zero-modes turn into  degenerate threshold modes, exactly as for monolayer graphene.

In the special case of a 
constant\footnote{For a 
constant magnetic field $B$ the assumption of compact support is not fulfilled.
 However, in this case the solution to $\partial_\zc W=-iA_+$ and   $\partial_z W=iA_-$  (\ie  $\partial_x W = A_y$ and $\partial_y W=- A_x$) is trivial. For the symmetric gauge $\vec{A}=\frac{1}{2}B(-y,x)$, for example,  one finds $W=\frac{1}{4}B (x^2+y^2)=Bz^*z$.
 The appearance the gaussian factor can also be extracted from \eref{wb} by considering $|\vec{x}|\ll R $ for$B(\vec{x}) = B_0\Theta (R-|\vec{x}|)$.}
 negative magnetic field $B$ the subspace of zero-modes is nothing but  the $J$ first Landau levels with $\Pi_+ /\sqrt{-2B}$ being a lowering operator of the Landau level index. The Dirac Landau levels with energy $E \propto \pm \sqrt{j}\neq 0 $ are given by 
$
(u,\, v)^\textrm{T}=(\varphi_j^l, \, \pm\varphi_{j-J}^l)^\textrm{T}
$
 with $j\geq J$ and in terms of the  Landau level $j$ wave-functions  $\phi_j^l(z,\zc)$ of a spinless \sch hamiltonian.  The zero-modes are given by $(\varphi_j^l,\, 0 )^\textrm{T}$ with $j=0,\ldots, J-1 $.   In particular, the zero energy LL of a bilayer has twice the degeneracy of the corresponding monolayer analog.\cite{MccFal} 

\section{A supersymmetric spectrum}\label{s:sym}
\noindent
In addition to the note on threshold modes in multilayers, we will also make a note on the rest of the spectrum. The chiral symmetry
\be{ph}
\sigma_z H_{m,A_0,\vec{A}}\sigma_z =- H_{ -m,-A_0,\vec{A}}
\ee
within each valley and valid for all $J$ implies that the symmetry of the spectrum  around $E=0$  is broken by a scalar potential or a mass term.  (For the massive case and $J$ odd, the particle-hole conjugation
$\sigma_x H^*_{m,A_\mu}\sigma_x =- H_{m,-A_\mu}$  guarantees a symmetric spectrum per valley, if instead gauge potentials are absent. For $J$ even, let $\sigma_x\rightarrow \sigma_y$.)
In particular, there are threshold modes  either only  at $E=+m$ or only at $E=-m$, depending on the sign of the total flux. 
 
 Amazingly, however, the rest of the spectrum within each valley remains symmetric, even when $m\neq 0$, provided the mass term $m$ is constant and the scalar potential $A_0$ is zero.  With these conditions the property
\be{susyconj}
0=[H^2,\sigma_z]=\{H, \frac{1}{2}[H,\sigma_z] \}\, ,
\ee
is fulfilled, 
which we note holds for any $J$ and arbitrary $\vec{A}$. It 
implies that the hermitian operator 
\be{q2}
\frac{i}{2}[H,\sigma_z]
=i\mtrx{cc}{0 & - \Pi_- \\ \Pi_+ & 0}
\ee
 maps positive energy states into negative energy states, except the threshold modes, which it kills. In the massless limit it coincides with the chiral symmetry \eref{ph}, except for zero-modes.  
   
To be more fancy, the symmetry is a manifestation of supersymmetry as defined in supersymmetric quantum mechanics.\cite{Wit} The Dirac hamiltonian is one important example\cite{CroRit}, but we can obviously generalize to the multilayer hamiltonians.
This has already been noted\cite{Eza}  in the study of the quantum Hall spectrum in multilayers.  However, the quantum Hall spectrum is highly degenerate, and we want to stress that the supersymmetry in multilayers holds for arbitrary magnetic fields and that the minimum message of  supersymmetry is most clearly seen in random fields.   
The real usefulness of redressing \eref{susyconj} into supersymmetry would be if one could make contact with the  rich literature on analytic results based on the latter. (Supersymmetry, for instance,  explains why the Dirac equation for some potentials, in particular 3d Dirac electrons in a Coulomb potential,  can be solved exactly and the spectrum can be constructed algebraically.) One thing that might be interesting for the study of ripples could be results on  scattering of Dirac particles in  slowly decreasing magnetic fields when asymptotic states are not easy to define, see \eg ref.~\onlinecite{Tha}.  For results relying on supersymmetry, we expect that similar results should hold for the supersymmetric hamiltonians of multilayer graphene.   

The essential structure is that there is a unitary self-adjoint operator $\tau$ (in our case $\sigma_z$) with $\tau^2=1$ for which there is a decomposition of $H$ in hermitian parts  $H=H_\textrm{odd}+H_\textrm{even}$ such that $[H_\textrm{even},\tau]=\{H_\textrm{odd},\tau\}=\{H_\textrm{even}, H_\textrm{odd}\}=0$.   
In such a case one can go on to form the supersymmetric hamiltonian $\Hss=\frac{1}{2}H_\textrm{odd}^2=2\Qss_1^2=2\Qss_2^2=\{\Qss,\Qss^\dagger\} $ and the supercharges $\Qss_1=\frac{1}{2}H_\textrm{odd}$, $\Qss_2=i[\Qss_1,\tau]$  and $\Qss=\frac{1}{2}(\Qss_1+i\Qss_2)$ with $\{\Qss_1,\Qss_2\}=0$ and $\Qss^2=0$. We recognize $2\Qss_1$ as the massless Dirac hamiltonian and $\Qss_2$ as the  operator \eref{q2} conjugating the spectrum of the massive Dirac hamiltonian. $\Qss$ and $\Qss^\dagger$ are fermionic ladder operators to be discussed below.  
For the hamiltonian  $H=D_x(\vec{x})\sigma_x+D_y(\vec{x})\sigma_y +m\sigma_z$  (for multilayers  $D_\pm =\Pi_\pm^J$) one finds
\be{}
H^2=\mtrx{cc}
{D_-D_+ & 0 \\ 0 & D_+D_-}+m^2 = 2\Hss +m^2\, .
\ee
 A supersymmetric hamiltonian  $\Hss$  is obviously positive definite. Because of $[\Hss,\Qss]=0$, all positive energy eigenvalues $E_i$ of $\Hss$ correspond to a degenerate doublet
\be{}
\left \{
\mtrx{c}{u_i \\ 0} , \, \mtrx{c}{0 \\ v_i }\right \} \,,
\ee
 one state coined "bosonic" and the other "fermionic". The  supercharges 
\be{}
\Qss=\mtrx{cc}{0 & D_-\\ 0 & 0}  \hspace{0.2cm} \textrm{and}
\hspace{0.2cm}
\Qss^\dagger= \mtrx{cc}{0 & 0\\ D_+ & 0}
\ee         
act as fermionic ladder operators stepping between them. The two states in one doublet are linear combinations of two eigenstates of $H$ related by $\Qss_2$. (In particular, in the case of constant $B$ the doublets are  $\{ (\varphi_j^l,\, 0 )^\textrm{T}, \, (0,\, \varphi_{j-J}^l)^\textrm{T}\}$.)

 The exceptions are the zero-modes of $\Hss$, also zero-modes of all the supercharges. (Thus, the zero-modes of $H_{m=0}$ or the threshold modes of $H$.)  Supersymmetry does not imply their existence. If they exist they do not need to come in pairs since they are killed  both by $\Qss$ and  by $\Qss^\dagger$. (Equivalently,  the threshold modes of $H$ cannot be conjugated by $\Qss_2$.) 
 
 The asymmetry in number of bosonic and fermionic zero-modes is the Witten index, which equals the Atiyah-Singer index for  the operators $D_\pm$ when they have the Fredholm property. (See \eg ref.~\onlinecite{Tha}.) For Dirac operators on an infinite plane, the Witten index defined as the asymmetry between bosonic and fermionic states should equal the number of zero-modes given by the Aharonov and Casher argument. However, with the analytical definition of the Witten index discussed in ref.~\onlinecite{Tha}, it actually remains $\Phi/\phi_0$ also on  the infinite plane.  The Witten index is then only related but not identical to the number of zero-modes according to Aharonov and Casher.

 \section{Summary} 
 \noindent
In this paper we pointed out that Aharonov-Casher argument for zero-modes for 2d Dirac electrons in a magnetic field generalizes naturally to the low-energy hamiltonians studied in the case of multilayer graphene. We also discussed the relationship of this work to that of Katsnelson and Prokhorova\cite{KatPro}, who gave a formal proof for bilayers based on the Atiyah-Singer index theorem. We extended the discussion to the presence of a uniform gap, in which case the degeneracy remains and might still be observable as a peak in the density of states, but at nonzero energy with the zero-modes instead being threshold modes.  

Further, we made a note on the symmetry of the spectrum within each valley in the presence of an arbitrary magnetic field. In the massless case the chiral symmetry guarantees a symmetric spectrum. Apart from the asymmetry of the threshold modes, the multilayer spectrum remains symmetric for a uniform nonzero mass term. We related this  to the  supersymmetric structure of the considered multilayer hamiltonians. We expect  that many of the analytic results based on the supersymmetry of the Dirac hamiltonian  should have analogs applying to the hamiltonians of multilayer graphene. Interesting applications might be found for example for the study of scattering in a nonuniform magnetic background.  

All these arguments fail if the scalar potential is not negligible compared to the vector potential. We brought up this issue but do not know of any solid answers neither for single-layer nor for multilayer graphene. 
  
 
{\it Acknowledgments}  The research was funded by the Swedish Research Council and a visiting grant of the Max-Planck Institut f\"ur Physik komplexer Systeme. The author is grateful to E. Mariani for discussions about the size of the scalar potential.

 \begin{center}
 {\bf Appendix: Aharonov's and Casher's argument}
 \end{center}
 \noindent
The two-dimensional Dirac equation $0=(H-E)\psi$  for electrons of mass $m$  and charge $-1$ is determined by the hamiltonian $H=\Pi_x\sigma_x+\Pi_y\sigma_y +m\sigma_z$.
 ($c=\hbar= e=1$, hence $\phi_0= h/e=2\pi$.)  The  threshold modes at $E=+m$ 
have to satisfy $\psi_{+m} = (u,\,  0 )^\mathrm{T}$ and $\Pi_+ u  = 0$. 
Likewise, there are  threshold modes $E=-m$ of the form  $\psi_{-m}=(0,v)^\mathrm{T}$ provided $\Pi_ -   v=0$. It will now be investigated which of these possibilities really gives normalizable wave functions.
Introduce $z= (x+iy)/2$, \ie  $\nabla^2 =\partial_z \partial_\zc $  and $\Pi_+=(-i \partial_\zc+A _+)$ with $A_\pm =  A_x \pm iA_y $. The Ansatz  $
u(\vec{x})=f(z)e^{ W(\vec{x})} $,
with $f(z)$ an arbitrary analytic function, satisfies $
0=\Pi_+ u =fe^W (-i\partial_\zc W +A_+)$. Acting with $\partial_z$ on  $\partial_\zc W =-iA_+$ gives the Poisson equation $\nabla^2 W= B-i\nabla \cdot \vec{A} $. Thanks to the boundary conditions given by the compact support of $B$ 
this equation can be inverted with the help of  $\nabla_{\vec{x}}^2 \ln |\vec{x}-\vec{x^\prime}| =2\pi \delta (\vec{x}-\vec{x^\prime})$, resulting in 
\be{wb}
\begin{array}{rl}
 W(\vec{x}) &= F(z)+G(\zc)+ \\
&+ \int \frac{\der^2 x^\prime}{2\pi}\,\, 
(B(\vec{ x^\prime})-i\nabla\cdot \vec{A}(\vec{x^\prime)} ) \ln |\vec{x}-\vec{x^\prime}|\, ,
\end{array} 
\ee
with $F(z)$ and $G(\zc)$ arbitrary analytic functions.
We put $F(z)=0$ as it is already accounted for by $f(z)$. Also, $\partial_\zc W=-iA_+$ implies $G(\zc)=0$.  Similarly, the Ansatz  $v(\vec{x})=f(\zc)e^{-W^*(\vec{x})}$ results in the same $W$ as for $u$.  
Choosing Coulomb gauge $\nabla \cdot \vec{A}=0$ implies  $W^*=W=\frac{1}{\phi_0}\int \der^2 x^\prime B(\vec{ x^\prime}) \ln |\vec{x}-\vec{x^\prime}| $. 

Far away from region of flux (\ie $|\vec{x}| >  |\vec{x^\prime}|$ and  $ \ln |\vec{x}-\vec{x^\prime}|\sim \ln |\vec{x} | $) one has asymptotically $e^{\pm W(\vec{x})}\sim |\vec{x}|^{\pm\Phi/2\pi}$. Normalizability imposes that only $E=+m$  ($E=-m$) threshold modes  can exist for $\Phi<0$ ($\Phi>0$).
The degeneracy of these modes comes from the possible choices of  $f(z)$.
 Single-valuedness of the wave function requires $f(z)=\sum_{s\in\mathbb{Z}} a_s z^s$  and normalizability at $|\vec{x}|\rightarrow \infty$   requires $\deg f < |\Phi|/\phi_0-1=n+\epsilon-1$, where $0<\epsilon \leq 1$.  Normalizability within the region of flux requires $B(\vec{x})$ to be \textit{non-singular} and all  powers in $f$ to be non-negative. Thus, $f$ is a polynomial in $z$ (or $\zc$) of maximal degree $n-1$. There are $n$ linearly independent polynomial  $\textrm{deg} \, f\leq n-1$, therefore the $n$-dimensional subspace of threshold modes.



\begin{thebibliography} {99}

{\footnotesize


\bibitem{Meyer} J. C. Meyer {\it et al.}, Nature {\bf 446}, 60-63 (2007).

\bibitem{Scott} J. Scott Bunch {\it et al.}, Science {\bf 315}, 490 (2007).

\bibitem{KanMel} C. L. Kane and E. J. Mele, Phys. Rev. Lett {\bf 78}, 1932 (1997).




\bibitem{Cas} A. H. Castro Neto {\it et al.}, Rev. Mod. Phys. {\bf 81}, 109 (2009).

\bibitem{GuiKatVoz} F. Guinea, M. I. Katsnelson and M. A. H. Vozmediano,
Phys. Rev. B {\bf 77}, 075422 (2008). 
\bibitem{Weh} T. O. Wehlin {\it al.}, 
Europhys. Lett. {\bf 84}, 17003 (2008).



\bibitem{Gie} A. J. Giesbers {\it et al.}, Phys. Rev. Lett. {\bf 99}, 206803 (2007).


\bibitem{Nak} M. Nakahara, {\it Geometry, Topolgy and Physics}, IOP Publishing (1990).


\bibitem{NieSem} A. J. Niemi and G. W. Semenoff, Phys. Rep. {\bf 135}, 99 (1986).   

\bibitem{Tha} B. Thaller, {\it The Dirac equation}, Springer Verlag Berlin Heidelberg (1992). 

\bibitem{AtiSin} M. F. Atiyah and I. M. Singer, Ann. Math. {\bf 87}, 485 and 546 (1968);    M. F. Atiyah and G. B. Segal, Ann. Math. {\bf 87}, 531(1968);
M. F. Atiyah, V. Patodi and I. M. Singer, Math. Proc. Cambridge. Phil. Soc. {\bf 77}, 43 and 405 (1975); {\bf 79}, 71 (1976).  



\bibitem{JacReb} R. Jackiw and C. Rebbi, Phys. Rev. D {\bf 13}, 3398 (1976).  

\bibitem{AhaCas} Y. Aharonov and A. Casher, Phys. Rev. A {\bf 19}, 246 (1979). 

\bibitem{KatPro} M. I. Katsnelson and M. F. Prokhorova,  Phys. Rev. B {\bf 77}, 205424 (2008). 

\bibitem{Eli} D. C. Elias {\it et al.}, arXiv:0810.4706. 

\bibitem{Tai} O. Taisuke {\it et al.}, Science {\bf 18}, Vol. 313, 951  (2006). 

\bibitem{Oos} J. B. Oostinga {\it et al.}, Nature Materials {\bf 7}, 151 (2008). 

\bibitem{Gie2009} A. J. M. Giesbers {\it et al.}, arXiv:0904.0948. 



\bibitem{Wit} E. Witten, Nucl. Phys. B {\bf 188}, 513 (1981); Nucl. Phys. B {\bf 202}, 253 (1982). 

\bibitem{Zha} Y. Zhang {\it et al.}, Phys. Rev. Lett. {\bf 96}, 136806 (2006). 

\bibitem{Aba} D. A. Abanin {\it et al.}, Phys. Rev. Lett. {\bf 98}, 196806 (2007).   


\bibitem{SuzAnd} H. Suzuura and T. Ando, Phys. Rev. B {\bf 65}, 235412 (2002).

\bibitem{And} T. Ando, J. Phys. Soc. Jpn. {\bf 75}, 124701 (2006).



\bibitem{MccFal} E. McCann and V. I. Fal'ko, Phys. Rev. Lett. {\bf 96}, 086805 (2006). 

\bibitem{GuiCasPer} F. Guinea, A. H. Castro and N. M. R. Peres, Phys. Rev. B {\bf 73}, 245426 (2006).

\bibitem{KosAnd} M. Koshino and T. Ando, Phys. Rev. B {\bf 76}, 085425 (2007).

\bibitem{MinMac} H. Min and A. H. MacDonald, Phys. Rev. B {\bf 77}, 155416 (2008).



\bibitem{MeyGei} J. C. Meyer {\it et al.},
Solid State Commun. {\bf 143}, 101 (2007).


\bibitem{Pal} R.S. Palais, with contributions by M. F. Atiyah, A. Borel, E. E. Floyd, R. T. Seeley, W. Shih, and R.. Solovay, {\it Seminar on the Atiyah-Singer Index Theorem}, Princeton University Press, NJ (1965).



\bibitem{CroRit} M. de Crombrugghe and V. Rittenberg, Ann. Phys. {\bf 151}, 99 (1983). 


\bibitem{Eza} M. Ezawa, Physica E {\bf 40}, 269 (2007); Phys. Lett. A {\bf 372}, 924 (2008). 




}
\end{thebibliography}
\end{document}